\documentclass{article}
\usepackage[%
journal=IFB,    
lang=british,   
]{ems-journal}

\DeclareMathOperator{\tr}{tr} 
\numberwithin{equation}{section}

\begin{document}

\title{A numerical approach for the dynamics of active viscoelastic surfaces}



\emsauthor{1}{
	\givenname{Francine}
	\surname{Kolley-Köchel}
	\orcid{0009-0008-2265-2305}}{F.~Kolley-Köchel}
\emsauthor{2}{
	\givenname{Jan Magnus}
	\surname{Sischka}
	\orcid{0009-0006-3154-2146}}{J.~M.~Sischka}
\emsauthor{3}{
	\givenname{Axel}
	\surname{Voigt}
	\orcid{0000-0003-2564-3697}}{A.~Voigt}
 \emsauthor{4}{
	\givenname{Elisabeth}
	\surname{Fischer-Friedrich}
	\orcid{}}{E.~Fischer-Friedrich}
\emsauthor{5}{
	\givenname{Sebastian}
	\surname{Aland}
	\orcid{}}{S.~Aland}

\Emsaffil{1}{
	\department{1}{Institute of Numerical Mathematics and Optimization}
	\organisation{1}{TU Bergakademie Freiberg}
	\address{1}{Akademiestraße 6}
	\zip{1}{09599}
	\city{1}{Freiberg}
	\country{1}{Germany}
    \affemail{francine.kolley-koechel@tu-dresden.de}
	\department{2}{Faculty of Informatics/Mathematics}
	\organisation{2}{HTW Dresden - University of Applied Sciences}
	\address{2}{Friedrich-List-Platz 1}
	\zip{2}{01069}
	\city{2}{Dresden}
	\country{2}{Germany}
	}
\Emsaffil{2}{
	\department{1}{Institute of Scientific Computing}
	\organisation{1}{Technische Universität Dresden}
	\zip{1}{01062}
	\city{1}{Dresden}
	\country{1}{Germany}
    \affemail{jan$\_$magnus.sischka@tu-dresden.de}
    }

\Emsaffil{3}{
	\department{1}{Institute of Scientific Computing}
	\organisation{1}{Technische Universität Dresden}
	\zip{1}{01062}
	\city{1}{Dresden}
	\country{1}{Germany}
	\affemail{axel.voigt@tu-dresden.de}
    \department{2}{Cluster of Excellence Physics of Life (PoL)}
    \organisation{2}{Technische Universität Dresden}
	\zip{2}{01062}
	\city{2}{Dresden}
	\country{2}{Germany}
    \organisation{3}{Center for Systems Biology Dresden (CSBD)}
    \address{3}{Pfotenhauerstraße 108}
	\zip{3}{01307}
	\city{3}{Dresden}
	\country{3}{Germany}
}

\Emsaffil{4}{
	\department{1}{Cluster of Excellence Physics of Life (PoL)}
	\organisation{1}{Technische Universität Dresden}
	\zip{1}{01062}
	\city{1}{Dresden}
	\country{1}{Germany}
	\affemail{elisabeth.fischer-friedrich@tu-dresden.de}
	}
    
\Emsaffil{5}{
	\department{1}{Institute of Numerical Mathematics and Optimization}
	\organisation{1}{TU Bergakademie Freiberg}
	\address{1}{Akademiestraße 6}
	\zip{1}{09599}
	\city{1}{Freiberg}
	\country{1}{Germany}
	\affemail{sebastian.aland@math.tu-freiberg.de}
	\department{2}{Faculty of Informatics/Mathematics}
	\organisation{2}{HTW Dresden - University of Applied Sciences}
	\address{2}{Friedrich-List-Platz 1}
	\zip{2}{01069}
	\city{2}{Dresden}
	\country{2}{Germany}
	\organisation{3}{Center for Systems Biology Dresden (CSBD)}
	\address{3}{Pfotenhauerstr. 108}
	\zip{3}{01307}
	\city{3}{Dresden}
	\country{3}{Germany}
	}
\classification[76A10,76M10]{92C05}


\begin{abstract}
The dynamics of active viscoelastic surfaces plays an important role in biological systems. One prominent example is the actin cortex, a thin bio-polymer sheet underneath the outer membrane of biological cells which combines active molecular force generation with viscoelastic behavior characterized by elastic properties at short timescales and viscous properties at longer timescales. We consider a surface Maxwell model within dominant rheology and an additional active term to model the dynamics of the actin cortex. This captures both, shear and dilational surface dynamics. We propose a monolithic numerical approach based on the surface finite element method (SFEM), validate the results for special cases and experimentally demonstrate convergence properties. Moreover, imposing a ring-shaped region of an enhanced active stress mimicking the contractile ring during cytokinesis, we observe different types of emergent patterns and shape dynamics depending on the viscoelastic properties. While viscous surfaces show a ring, which slips to one side of the surface, viscoelasticity provides a stabilization mechanism of the ring, thus forming a requirement  for subsequent cell division. This study provides an example that viscoelastic properties are key ingredients to understand biological materials. 
\end{abstract}

\maketitle


\section{Introduction}
\label{sec:Introduction}
The dynamics of active surfaces are central to understanding complex biological processes such as tissue morphogenesis, cell membrane remodeling, and cell shape changes \cite{Taubenberger2020, Ferreira2020, Magidson2011, Flanary2019}. A prime example of such an active surface is the cell cortex — a thin, viscoelastic layer lining the inner face of the cell membrane in eukaryotic cells \cite{Alberts}. 
During cytokinesis, the bio-polymers in the cortex become more concentrated at the cell equator forming a contractile ring. This ring is ultimately dividing the cell into two daughter cells. This process is a key step in the cellular reproduction cycle and therefore fundamental to growth and proliferation in living organisms. While the molecular architecture of the key components involved in this process are well understood, the physical principles still remain under debate.

Various models have been proposed to at least qualitatively describe the essential phenomena \cite{Torres2019, da2022viscous, al2021active, Nitschke2025}. These are highly nonlinear surface models combining flows, forces, internal degrees and geometric properties and even minimal models, which are isotropic and only consider an active isotropic force, exhibit strong nonlinearities. This active isotropic force is tied to surface tension, influencing the energy reaction to changes in area \cite{al2023, Nitschke2025}. Within various approaches this active isotropic force has been used to model mechanical feedback by considering the strength as a function of a stress regulator molecule, see \cite{Bois2011} for the model in one dimension and \cite{Wittwer2023, Mietke2019} for numerical realizations in an axisymmetric setting. They all consider viscous surfaces, see also \cite{Bonati2022, neiva2025unfitted}, and in many of these studies the self-organized formation of a contractile ring has been observed \cite{Wittwer2023, Mietke2019, Bonati2022}. Models for viscoelastic surfaces, without or with active forces, are less explored. Correspondingly, the influence of viscoelasticity on the formation of a contractile ring and the onset of cell division remains still elusive. 

We here follow a proposed model for a viscoelastic active surface considered as an interface conditions for surrounding viscous fluids \cite{Kinkelder2021, Kinkelder2023}. While this setup simplifies and stabilizes the numerics, it becomes unstable when surface viscosity dominates—and fails to accurately reflect systems like the actin cortex, where surface rheology is highly dominant \cite{Fischer2016}. Moreover, embedding the surface in a surrounding fluid greatly increases computational cost, expanding an essentially two-dimensional problem into a fully three-dimensional domain. To overcome these limitations we consider the active viscoelastic surface in the absence of surrounding fluids and propose a numerical framework for its solution. 
In a range of numerical tests, we validate the model and show that it performs well in the limits of purely viscous and purely elastic surfaces and demonstrate convergence. 
Finally, we use the model to get a first glimpse on the self-organized patterns of active viscoelastic surfaces. The active isotropic force is thereby treated as in the viscose case. Following \cite{Mietke2019} we introduce a scalar function describing stress regulator concentration on the surface. We show that viscoelasticity can stabilize the ring-shaped concentration pattern, which is a requirement for successful cell division. Studying biologically relevant parameters, we discuss how this model can provide insights into cytokinesis and asymmetric cell division. 

\section{Modelling of active viscoelastic surfaces}
\label{sec:Modelling}

For the actin cortex the surface is the dominant mechanical element and contributions by a surrounding fluid are negligible. Therefore, we restrict the whole computational domain to an evolving surface $\Gamma \subset \mathbb{R}^3$. We assume it to be closed without boundary and sufficiently regular. The surface projection operator is defined as $\mathbf{P} = \mathbf{I} - \boldsymbol{n} \boldsymbol{n}^T$, where $\boldsymbol{n} : \Gamma \rightarrow \mathbb{R}^3$ is the (inward pointing) unit normal vector of the surface $\Gamma$ and $\mathbf{I}$ the identity in $\mathbb{R}^3$. 
Let $c$ be a scalar field, $\boldsymbol{v}$ a $\mathbb{R}^3$ vector field and $\mathbf{D}$ a $\mathbb{R}^{3\times3}$ tensor field on $\Gamma$. Let $c$, $\boldsymbol{v}$ and $\mathbf{D}$ be continuously differentiable with arbitrary smooth extensions $c^e$, $\boldsymbol{v}^e$ and $\mathbf{D}^e$ in normal direction. We define the following surface gradients $\nabla_\Gamma c = \mathbf{P} \nabla c^e$, $\nabla_C \boldsymbol{v} = \nabla\boldsymbol{v}^e \mathbf{P}$, and $\nabla_C\mathbf{D} = \nabla\mathbf{D}^e\mathbf{P}$, where $\nabla$ is the gradient of the embedding space $\mathbb{R}^3$. The corresponding divergence operators are $\operatorname{div}_{C} \boldsymbol{v} = \operatorname{tr}(\nabla_C\boldsymbol{v})$ and $\operatorname{div}_C(\mathbf{DP}) = \operatorname{tr}\nabla_C(\mathbf{D}\mathbf{P})$, where $\operatorname{tr}$ is the trace operator. Note that
${\operatorname{div}_C\boldsymbol{v} = \operatorname{div}_\Gamma(\mathbf{P}\boldsymbol{v})-(\boldsymbol{v}\cdot\boldsymbol{n})H}$, where $\operatorname{div}_\Gamma$ is the covariant divergence defined for tangential vector fields and $H$ denotes the mean curvature of $\Gamma$.
 

The surface is modeled as a viscoelastic Maxwell material, with an elastic response to deformations on short time scales and viscous resistance on larger time scales. The model is adapted from \cite{Kinkelder2021, Kinkelder2023}.
Let $\boldsymbol{v} : \Gamma \rightarrow \mathbb{R}^3$ denote the velocity field of the surface, comprising both tangential and normal components. The viscous response depends on the strain rate tensor $\mathbf{D}:\Gamma \rightarrow \mathbb{R}^{3\times 3}$ given by 
\begin{equation} \label{eq:deformation rate tensor}
    \mathbf{D} = \frac{1}{2} \left( \mathbf{P}\nabla_C \boldsymbol{v} + (\mathbf{P}\nabla_C \boldsymbol{v})^T   \right) ~.
\end{equation}
with its traceless component 
\begin{equation}
    \bar{\mathbf{D}}= \mathbf{D} - \frac{1}{2} \tr{\mathbf{D}}~\mathbf{P},
\end{equation}
where $\tr{\mathbf{D}}=\operatorname{div}_C\boldsymbol{v}$.

The force balance of the hydrodynamic system is given by
\begin{equation}
\rho_\Gamma (\partial_t \boldsymbol{v} + \nabla_{\boldsymbol{w}} \boldsymbol{v} ) + q \boldsymbol{n} = \operatorname{div}_C (\mathbf{S} + \xi f(c) \mathbf{P} ) ~,
\label{eq:force}
\end{equation}
where  $[\nabla_{\boldsymbol{w}}\boldsymbol{v}]_i = \nabla_\Gamma \boldsymbol{v}_i \cdot\boldsymbol{w}$, $i = 1, 2, 3$, with $\boldsymbol{w} = \boldsymbol{v} - \partial_t \mathbf{x}$ the relative material velocity,  $\mathbf{x}$ the parametrization of $\Gamma$, and $\partial_t$ the time derivative along this parametrization, capturing the evolution at fixed reference coordinates. The first term describes inertial surface forces with surface mass density $\rho_\Gamma$. Despite its minimal influence on the dynamics at low Reynolds number, this term ensures a well-defined solution.
The second term, $q\boldsymbol{n}$, represents the pressure contribution of the fluid encapsulated by $\Gamma$, which is not explicitly modeled. Thereby, $q$ acts as a Lagrange multiplier, which ensures $\int_\Gamma \boldsymbol{v} \cdot \mathbf{n} ~{\textrm d}\Gamma= 0$, see also \cite{krause2023numerical,Kinkelder2025}.
The right-hand side contains the surface stresses, with the viscoelastic surface stress $\mathbf{S}$ and the active stress $\xi f(c) \mathbf{P}$, scaled with a constant $\xi$ and regulated by a surface field $c:\Gamma\rightarrow\mathbb{R}$ which describes the concentration of force-generating molecules. This part thus results in $\operatorname{div}_C (\xi f(c) \mathbf{P}) = \xi (f^\prime(c) \nabla_\Gamma c - f(c) H \boldsymbol{n})$, with $H = -\operatorname{div}_C \boldsymbol{n}$ the mean curvature. As in previous literature \cite{Kinkelder2021, Kinkelder2023, Wittwer2023, Bonati2022, Mietke2019}, we use a Hill-function
\begin{equation}
f(c)=\frac{2c^2}{c_0^2+c^2}
\end{equation}
and determine the concentration field by a surface convection-diffusion equation,
\begin{equation}
\partial_t c + \nabla_{\boldsymbol{w}}  c + ( \operatorname{div}_C\boldsymbol{v}) c = D_\mathrm{c}\Delta_\Gamma c - k_\mathrm{off} c + k_\mathrm{on} c_0,
\label{eq:concentration}
\end{equation}
with diffusion constant $D_\mathrm{c}$. The change in concentration by convection is given by $\nabla_{\boldsymbol{w}} c$, and the change by surface shrinkage/growth is modeled via $(\operatorname{div}_C \boldsymbol{v}) c$. The rate $k_\mathrm{off}$ models unbinding of molecules from the surface. In addition, an autocatalytic binding rate $k_\mathrm{on}c_\mathrm{0}$ increases the surface concentration. 

The stress tensor $\mathbf{S}$ represents viscoelastic stresses acting on the surface and is divided into its traceless part $\bar{\mathbf{S}}$ and dilational part, represented by $\tr\mathbf{S}$:
\begin{equation}
\mathbf{S}=\bar{\mathbf{S}} +\frac{1}{2} \tr\mathbf{S}~\mathbf{P}.
\label{eq:stress}
\end{equation}
This decomposition accounts for the two modes of deformation of a surface: shear and dilation (see Fig. 1). Shear deformation changes the aspect ratio of surface elements, while their area remains constant. In a viscoelastic model, this mode of deformation affects the shear stress $\bar{\mathbf{S}}$, associated with shear viscosity $\eta_\mathrm{s}$ and the corresponding relaxation time $\tau_\mathrm{s}$. 
Surface dilation, on the other hand, changes the dilational stress $\tr{\mathbf{S}}$ as surface elements grow ($\tr{\mathbf{S}}>0$) or shrink ($\tr{\mathbf{S}}<0$), while their proportions remain constant. The evolution of $\tr\mathbf{S}$ is determined by the surface bulk viscosity $\eta_\mathrm{b}$ and the corresponding relaxation time $\tau_\mathrm{b}$.

The stresses evolve according to,
\begin{align}
\tr\mathbf{S} &= 2 \eta_\mathrm{b} \tr{\mathbf{D}} + \tau_\mathrm{b} \left(2(\bar{\mathbf{S}}:\nabla_C \boldsymbol{v}) + \tr\mathbf{S}\tr{\mathbf{D}} - \partial_t^\bullet \tr\mathbf{S} \right)~, \label{eq:dilational_stress_0} \\
\bar{\mathbf{S}} &= 2 \eta_\mathrm{s} \bar{\mathbf{D}} - \tau_\mathrm{s} \partial_t^\triangledown \bar{\mathbf{S}} ~, \label{eq:shear_stress_0}
\end{align}
with the material derivative $\partial_t^\bullet =\partial_t+\nabla_{\boldsymbol{w}}$ and traceless upper convected derivative 
\begin{equation} \label{eq:upper convected}
\partial_t^\triangledown {\bar{\mathbf{S}}} = \partial_t^\bullet \bar{\mathbf{S}} - \nabla_C \boldsymbol{v} \bar{\mathbf{S}} - \bar{\mathbf{S}} ( \nabla_C\boldsymbol{v} )^T + (\bar{\mathbf{S}}:\nabla_C \boldsymbol{v} )\mathbf{P} -\tr\mathbf{S}~ \bar{\mathbf{D}}  ~. 
\end{equation}
The latter derivative rotates and translates a traceless tensor with the flow, while stretching it with tangential flows and keeping it traceless.  

Equations~\eqref{eq:stress}–\eqref{eq:upper convected} characterize the surface as a viscoelastic Maxwell fluid. In the following, we briefly outline how these equations connect to classical bulk viscoelastic models.
To this end, we combine the shear and dilational stress equations by adding  
$\frac{1}{\tau_{\rm s}}$ times eq.~\eqref{eq:shear_stress_0} and  $\frac{1}{2\tau_{\rm b}}\mathbf{P}$ times eq.~\eqref{eq:dilational_stress_0} and using eq.~\eqref{eq:upper convected}, yielding 
\begin{equation}
{
\partial_t^\bullet \bar{\mathbf{S}} + \frac{1}{2}\mathbf{P}\partial_t^\bullet \tr \mathbf{S}-
\nabla_C\boldsymbol{v}~\bar{\mathbf{S}} - \bar{\mathbf{S}} ( \nabla_C \boldsymbol{v} )^T + \mathbf{D}\tr \mathbf{S}= 
\frac{2\eta_s}{\tau_s} \bar{\mathbf{D}} - \frac{1}{\tau_s}\bar{{\mathbf{S}}}
+\frac{\eta_b}{\tau_b} \mathbf{P}\tr {\mathbf{D}} - \frac{1}{2\tau_b}\mathbf{P}\tr \mathbf{S}
}.\label{eq:2.10}
\end{equation}
Using the identity $\partial_t^\bullet \mathbf{P} = \nabla_C\boldsymbol{v}+ ( \nabla_C \boldsymbol{v} )^T - 2\mathbf{D}$ as derived from Lemma 37 in \cite{barrett2020parametric} and eq.~\eqref{eq:stress}, we can deduce $\partial_t^\bullet \mathbf{S} = \partial_t^\bullet \bar{\mathbf{S}}+\frac{1}{2}\mathbf{P}\partial_t^\bullet \tr {\mathbf S} + \frac{1}{2} \tr \mathbf{S} (\nabla_C\boldsymbol{v}+ ( \nabla_C \boldsymbol{v} )^T - 2\mathbf{D})$.
Applying this identity to the left-hand side of eq.~\eqref{eq:2.10} and noting that $\nabla_C\boldsymbol{v}=\nabla_C\boldsymbol{v}\mathbf{P}$, we obtain
\begin{equation}
{
\partial_t^\bullet {\mathbf{S}} - \nabla_C\boldsymbol{v}~{\mathbf{S}} - {\mathbf{S}} ( \nabla_C \boldsymbol{v} )^T = 
\frac{2\eta_s}{\tau_s} \bar{\mathbf{D}} - \frac{1}{\tau_s}\bar{{\mathbf{S}}}
+\frac{\eta_b}{\tau_b} \mathbf{P}\tr {\mathbf{D}} - \frac{1}{2\tau_b}\mathbf{P}\tr \mathbf{S}
}. 
\end{equation}
If we assume the surface to be flat, we obtain
\begin{equation}
{
\partial_t^\bullet {\mathbf{S}}_{} - \nabla_{\rm 2D} \boldsymbol{v}~{\mathbf{S}}_{} - {\mathbf{S}}_{} ( \nabla_{\rm 2D}  \boldsymbol{v} )^T = 
\frac{2\eta_s}{\tau_s} \bar{\mathbf{D}}_{} - \frac{1}{\tau_s}\bar{{\mathbf{S}}}_{}
+\frac{\eta_b}{\tau_b} \mathbf{I}_{} \tr {\mathbf{D}}_{} - \frac{1}{2\tau_b} \mathbf{I}_{} \tr \mathbf{S}_{}
}, 
\end{equation}
with $\nabla_{\rm 2D}$ the gradient in $\mathbb{R}^2$, and
$\boldsymbol{v}$, $\boldsymbol{w}, \mathbf{S} = \bar{\mathbf{S}}+\frac{1}{2}\mathbf{I}\tr \mathbf{S}$, $\mathbf{D}=\frac{1}{2}(\nabla_{\rm 2D}\boldsymbol{v}+\nabla_{\rm 2D}\boldsymbol{v}^T) = \bar{\mathbf{D}}+\frac{1}{2}\mathbf{I}\tr \mathbf{D}$ are vector- and tensor-fields in $\mathbb{R}^2$ and $\mathbb{R}^{2\times2}$, respectively

This result can be coupled to the analogue of eq. \eqref{eq:force} in flat space, neglecting the pressure contribution and the active stress, which reads  
\begin{equation}
\rho (\partial_t \boldsymbol{v} + \nabla_{\boldsymbol{w}} \boldsymbol{v} ) = \operatorname{div}_{\rm 2D} \mathbf{S}_{} ~,
\end{equation}
with $\operatorname{div}_{\rm 2D}$ the divergence in $\mathbb{R}^2$, to yield the compressible Maxwell model in two dimensions \cite{truesdell2004non}.
 
\begin{center}
\begin{figure}
  \includegraphics[width = 0.85\textwidth]{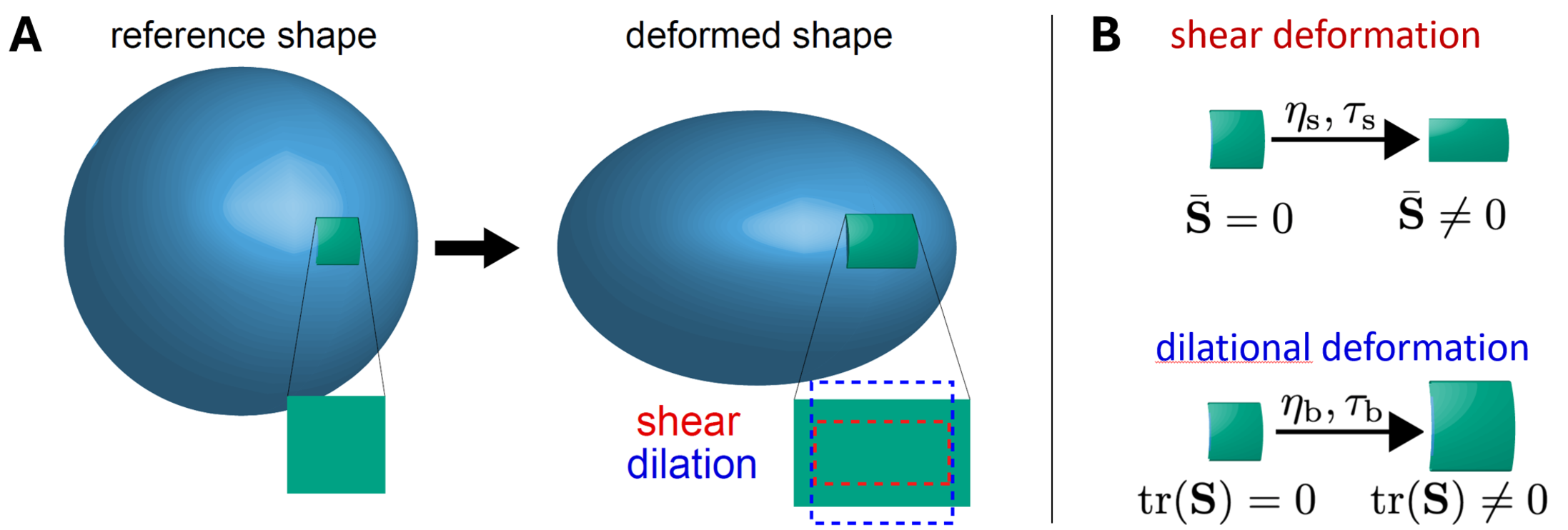} 
  \caption{Illustration of the two distinct modes of deformation with their relevant parameters. 
  \textbf{A.} Deformation of a square-shaped surface element (green). 
  \textbf{B.} Deformation can be decomposed into shear and dilation. 
  Shear stress $\bar{\mathbf{S}}$ measures changes in aspect ratio of the surface element and is  associated with surface shear viscosity $\eta_\mathrm{s}$ and relaxation time $\tau_\mathrm{s}$. 
  The dilational stress $\tr{\mathbf{S}}$ changes as the surface area grows ($\tr{\mathbf{S}}>0$) or shrinks ($\tr{\mathbf{S}}<0$) isotropically and is associated with the surface bulk viscosity $\eta_\mathrm{b}$ and  relaxation time $\tau_\mathrm{b}$.
  } 
  \label{fig:sketches}
\end{figure}
\end{center}

\section{Weak formulation and numerical implementation}
\label{sec:Numerical_implementation}

We now introduce a finite element discretization for the full system of equations~\eqref{eq:deformation rate tensor}–\eqref{eq:upper convected}, which comprises a coupled set of scalar-, vector-, and tensor-valued surface PDEs. To facilitate implementation, we adopt a component-wise solution strategy. This results in a total of 14 unknowns for $c, \boldsymbol{v}, \tr \mathbf{S}$, and $\bar{\mathbf{S}}$. 
Although $\bar{\mathbf{S}}$ is symmetric by construction, this symmetry is not exploited in the discretization, allowing for a uniform treatment of all tensor components.

The discretization considers a surface finite element method (SFEM) \cite{dziuk2013finite,nestlerFiniteElementApproach2019} in space and a semi-implicit finite difference method in time within an Arbitrary Lagrangian-Eulerian (ALE) approach \cite{elliottALEESFEMSolving2012}, primarily Lagrangian in normal direction and Eulerian in tangential direction. This requires a parameterization of $\Gamma$ and we follow the approach of \cite{barrettParametricApproximationWillmore2008} and include the contracted Gauss-Weingarten equation as an auxiliary equation
\begin{align}
  \label{eq::structuralEquation}
\boldsymbol{\kappa} &= \Delta_C\mathbf{x}
\\
\frac{d}{dt}\mathbf{x}&= \boldsymbol{v} \cdot \mathbf{n} \mathbf{n}.\label{eq::structuralEquation2}
\end{align}
with the parameterization $\mathbf{x}$, where $\boldsymbol{\kappa} = H\boldsymbol{n}$, which also provides a way to obtain $H$. 

We consider a discrete approximation $\Gamma_h$ of $\Gamma$, given by a triangulation with piecewise flat triangles, with $h$ the size of the mesh elements, i.e. the longest edge. We consider each geometrical quantity like the normal vector $\mathbf{n}_h$, and the inner products $(\cdot , \cdot)_h$ with respect to $\Gamma_h$. We define the discrete function spaces with $P^1$ elements.
We discretize in time using a semi-implicit time stepping scheme with constant step size $k$. In each time step we solve the full system of equations on the previous triangulation. 
Then, the mesh nodes are displaced by $k \boldsymbol{v}_N^i$, where $\boldsymbol{v}_N^i = (\boldsymbol{v}^i\cdot\boldsymbol{n}^i)\boldsymbol{n}^i$ is the normal part of the velocity field $\boldsymbol{v}$ to capture surface deformation, and the simulation proceeds to the next time step.

We consider a monolithic approach, assembling all equations in a single large system. However, we present the discretized version of the weak formulations of all involved equations separately. 

For the concentration field, we require for all suitable test functions $\Phi_c:\Gamma_h\rightarrow\mathbb{R}$, that

\begin{align}
0= &\int_{\Gamma_h} \nabla_\Gamma c_h^{i+1} \cdot \nabla_\Gamma \Phi_\mathrm{c} ~\mathrm{d}\Gamma 
- \int_{\Gamma_h} \left( \frac{c^i}{k} + k_\mathrm{on} c_0 \right)\Phi_\mathrm{c} ~\mathrm{d}\Gamma \notag \\
  +&\int_{\Gamma_h} \left( 
    \frac{c_h^{i+1}}{k} 
    + \nabla_{\boldsymbol{w}_h^i} c_h^{i+1} 
    + c_h^i \operatorname{div}_C \boldsymbol{v}_h^{i+1} 
    + k_\mathrm{off} c_h^{i+1} 
\right)\Phi_\mathrm{c} ~\mathrm{d}\Gamma ~.
\label{eq:c eq weak}
\end{align}


It remains to specify the force balance with the viscoelastic stress contributions. 
A numerical challenge in viscoelastic fluid simulations arises with high Weissenberg numbers, which represent the ratio between the fluid's relaxation time and the characteristic deformation time. This difficulty primarily stems from the advective nature of the stress evolution equation, which necessitates specialized stabilization techniques to prevent numerical oscillations. However, typical active viscoelastic surfaces in biology, such as the cell cortex, operate in low Weissenberg number regimes. Focusing on this regime enables us to proceed with a straightforward discretization without the need for explicit convection stabilization.

A semi-implicit discretization of the dilational stress is given by: Find $\tr\mathbf{S}_h^{i+1}$ such that for all suitable test functions $\Phi:\Gamma_h\rightarrow\mathbb{R}$ 
\begin{align}
0 &= \int_{\Gamma_h} \left(2\eta_\mathrm{b} \operatorname{div}_C \boldsymbol{v}_h^{i+1} + \tau_\mathrm{b} \tr\mathbf{S}_h^i \operatorname{div}_C \boldsymbol{v}_h^{i+1} + 2 \tau_\mathrm{b} (\bar{\mathbf{S}}_h^i:\nabla_C \boldsymbol{v}_h^{i+1}) \right) \Phi ~{\textrm d}\Gamma \notag \\
&+ \int_{\Gamma_h}\left(\frac{\tau_\mathrm{b}}{k} \tr{\mathbf{S}}_h^i -\left(1+\frac{\tau_\mathrm{b}}{k} \right)\tr\mathbf{S}_h^{i+1} - \tau_\mathrm{b}\nabla_{\boldsymbol{w}_h^i} \tr\mathbf{S}_h^{i+1} \right)\Phi ~{\textrm d}\Gamma,
\label{eq:dil_stress_weak}
\end{align}
and for the shear stress: Find $\bar{\mathbf{S}}_h^{i+1}$ such that for all suitable test function $\bar{\Phi}:\Gamma_h\rightarrow\mathbb{R}^{3\times 3}$
\begin{align}
    0 &= \int_{\Gamma_h} \left(\eta_\mathrm{s} + \frac{1}{2}\tau_\mathrm{s} \tr\mathbf{S}_h^i \right) \left(2\mathbf{D}_h^{i+1} - \nabla_C \boldsymbol{v}_h^{i+1} \mathbf{P}_h^i \right):\bar{\Phi} ~{\textrm d}\Gamma \notag \\
        &+ \int_{\Gamma_h}  \tau_\mathrm{s} \left( \nabla_C \boldsymbol{v}_h^{i+1} \bar{\mathbf{S}}_h^i + \bar{\mathbf{S}}_h^i ( \nabla_C \boldsymbol{v}_h^{i+1})^T -  (\bar{\mathbf{S}}_h^i:\nabla_C \boldsymbol{v}_h^{i+1} ) \mathbf{P}_h^i \right):\bar{\Phi}                      ~{\textrm d}\Gamma \notag \\
        &+ \int_{\Gamma_h} \left(\frac{\tau_\mathrm{s}}{k} \bar{\mathbf{S}}_h^i - \left(1+\frac{\tau_\mathrm{s}}{k} \right)\bar{\mathbf{S}}_h^{i+1} -\tau_\mathrm{s} \nabla_{\boldsymbol{w}_h^i} \bar{\mathbf{S}}_h^{i+1}\right):\bar{\Phi} ~{\textrm d}\Gamma.
        \label{eq:shear_stress_weak}
\end{align}
To complete the formulation, we need to  specify the weak form of the force balance equation \eqref{eq:force}. Given the quantities $\bar{\mathbf{S}}_h^{i+1}$ and $\tr\mathbf{S}_h^{i+1}$ from Eqs.~\eqref{eq:dil_stress_weak}-\eqref{eq:shear_stress_weak}, one could, in principle, reconstruct the full stress tensor via $\mathbf{S} = \bar{\mathbf{S}} + \frac{1}{2} (\tr\mathbf{S})\, \mathbf{P}$, and evaluate the viscoelastic force $\operatorname{div}_C \mathbf{S}$ entering the force balance equation.
However, we observed that this strategy leads to numerical instabilities and pronounced spatial oscillations. This behavior is, to some extent, expected, as it corresponds to approximating second-order spatial derivatives of $\boldsymbol{v}$ by two first-order equations.

To mitigate this issue, we define the viscoelastic force explicitly as $\mathbf{f} := \operatorname{div}_C\mathbf{S}$ and compute it by substituting the stress evolution equations \eqref{eq:dilational_stress_0} and \eqref{eq:shear_stress_0} directly into this expression. In doing so, $\mathbf{f}$ naturally incorporates second-order derivatives of the velocity field, allowing for a more stable and accurate discretization.
The discretized version of the weak formulation of the corresponding equation is: Find ${\textbf{f}}_h^{i+1}$ such that for all suitable test functions $\boldsymbol{u}:\Gamma_h\rightarrow\mathbb{R}^3$
\begin{align}
     \int_{\Gamma_h} {\textbf{f}}_h^{i+1}\cdot {\boldsymbol{u}} ~{\textrm d}\Gamma 
     &=\int_{\Gamma_h} \left(2\gamma_\mathrm{s} \left(\eta_\mathrm{s} + \frac{1}{2}\tau_\mathrm{s} \tr\mathbf{S}_h^i \right) \mathbf{D}_h^{i+1} \right) : \nabla_C{\boldsymbol{u}} ~{\textrm d}\Gamma \notag \\
     &+ \int_{\Gamma_h} \left( \left(\gamma_\mathrm{b} \eta_\mathrm{b} -\gamma_\mathrm{s} \eta_\mathrm{s} + \frac{1}{2}(\gamma_\mathrm{b} \tau_\mathrm{b} -\gamma_\mathrm{s} \tau_\mathrm{s}) \tr\mathbf{S}_h^i\right) \operatorname{div}_C \boldsymbol{v}_h^{i+1} \right) \mathbf{P}_h^i:\nabla_C{\boldsymbol{u}} ~{\textrm d}\Gamma \notag \\
    &+ \int_{\Gamma_h} (\gamma_\mathrm{b} \tau_\mathrm{b} -\gamma_\mathrm{s} \tau_\mathrm{s}) (\bar{\mathbf{S}}_h^i:\nabla_C \boldsymbol{v}_h^{i+1}) \mathbf{P}_h^i  :\nabla_C{\boldsymbol{u}} ~{\textrm d}\Gamma \notag \\
    &- \int_{\Gamma_h} \frac{1}{2}\tau_\mathrm{b}\gamma_\mathrm{b}  \left( \nabla_{\boldsymbol{w}_h^i} \tr\mathbf{S}_h^{i+1}  \right) \mathbf{P}_h^i: \nabla_C{\boldsymbol{u}} ~{\textrm d}\Gamma \notag \\
    &- \int_{\Gamma_h} \tau_\mathrm{s}\gamma_\mathrm{s}   \left( \nabla_{\boldsymbol{w}_h^i} \bar{\mathbf{S}}_h^{i+1} \right): \nabla_C{\boldsymbol{u}}  ~{\textrm d}\Gamma \notag \\
    &+ \int_{\Gamma_h} \left( \frac{\tau_\mathrm{b}\gamma_\mathrm{b}}{2k} \tr\mathbf{S}_h^i~ \mathbf{P}_h^i - \frac{\tau_\mathrm{s} \gamma_\mathrm{s}}{k} \bar{\mathbf{S}}_h^i \right) : \nabla_C{\boldsymbol{u}} ~{\textrm d}\Gamma \notag \\
    &+ \int_{\Gamma_h} \tau_\mathrm{s}  (\nabla_C \boldsymbol{v}_h^{i+1}\bar{\mathbf{S}}_h^i+ \bar{\mathbf{S}}_h^i(\nabla_C \boldsymbol{v}_h^{i+1})^T) : \nabla_C{\boldsymbol{u}} ~{\textrm d}\Gamma ~,
    \label{eq:fstress_weak}
\end{align}
where we have introduced the constants 
\begin{equation}
   \gamma_\mathrm{b}=\frac{1}{1+\frac{\tau_\mathrm{b}}{k}}   ~, \text{and}~\gamma_\mathrm{s}=\frac{1}{1+\frac{\tau_\mathrm{s}}{k}} ~.
\end{equation}
for convenience of notation. 
Note, that the first term on the right-hand side leads to a purely viscous term when included in the momentum equation (below), thereby adding an elliptic contribution. 
In the context of viscoelastic three-dimensional fluids, it has been demonstrated that the inclusion of a purely viscous term in the momentum equation ensures a stable finite element discretization \cite{baranger1992formulation}. However, a rigorous numerical analysis of this coupling for the present surface system, lies beyond the scope of the this work.

The discretized version of the weak formulation of the force balance is: Find $\boldsymbol{v}_h^{i+1}$ such that for all suitable test function $\boldsymbol{u}:\Gamma_h \rightarrow \mathbb{R}^3$
\begin{align}
0 &= \int_{\Gamma_h} \left(\rho_\Gamma \left(\frac{\boldsymbol{v}_h^{i+1}-\boldsymbol{v}_h^i}{k} + \nabla_{\boldsymbol{w}_h^i}\boldsymbol{v}^{i+1} \right) + q\boldsymbol{n}_h^i \right) \cdot \boldsymbol{u} ~{\textrm d}\Gamma  \notag \\
&- \int_{\Gamma_h} \xi \left(f'(c_h^i)\nabla_\Gamma c_h^{i+1} - f(c_h^i) \boldsymbol{\kappa}_h^{i+1} \right) \cdot \boldsymbol{u} ~{\textrm d}\Gamma \notag \\
&- \int_{\Gamma_h} {\textbf{f}}_h^{i+1} \cdot {\boldsymbol{u}} ~{\textrm d}\Gamma ~.
\label{eq:weak_force_balance}
\end{align}
The pressure $q$ is approximated by a penalty approach,
\begin{equation}
q=\alpha \frac{V_h^0-V_h^i}{V_h^0}~, \label{eq: q}
\end{equation} 
where $\alpha$ is a large constant with units of force density. This approach ensures that the volume $V_h^i$ enclosed by the surface at time step $i$ does not significantly deviate from its initial value $V_h^0$. 

Finally, the discretized version of Eqs. \eqref{eq::structuralEquation},
\eqref{eq::structuralEquation2} reads:
Find $(\boldsymbol{x}^{i+1}, \boldsymbol{\kappa}_h^{i+1})$ such that for all suitable test functions $\Phi_x,\Phi_\kappa:\Gamma\rightarrow\mathbb{R}^3$
\begin{align}
0 &=	 \int_{\Gamma_h} \boldsymbol{x}_h^{i+1} \cdot \Phi_x - \left( \boldsymbol{x}_h^i + k (\textbf{n}_h^i \cdot \boldsymbol{v}_h^{i+1}) \boldsymbol{n}_h^i \right) \cdot \Phi_x ~{\textrm d}\Gamma, \label{weak2} \\
0 &= \int_{\Gamma_h} \left( \boldsymbol{\kappa}_h^{i+1}  \cdot \Phi_\kappa + \nabla_C \boldsymbol{x}_h^{i+1} : \nabla_C \Phi_\kappa \right)~{\textrm d}\Gamma. \label{weak3}
\end{align}

The computed curvature vector $\boldsymbol{\kappa}$ points in the normal direction and can, in principle, be used to recover the surface normal vector. However, direct reconstruction of $\boldsymbol{n}$ from $\boldsymbol{\kappa}$ was found to be numerically unstable, particularly under large surface deformations where regions of low curvature may occur. To ensure robustness, we instead determine the normal vector $\boldsymbol{n}$ on each surface element via the cross product of edge vectors that span the element.

The numerical implementation of the model equations \eqref{eq:c eq weak}-\eqref{eq: q} is realized in C++ , using the finite element library AMDiS \cite{AMDiS,Witkowski2015} on top of the Dune framework \cite{DUNE}. The surface mesh was created with the mesh generator Gmsh \cite{Geuzaine2009}.
Simulations are conducted with the \textsc{CurvedGrid} module \cite{praetorius2020curvedgrid}.

\section{Results}
\label{sec:Results}
From here on, to improve readabilty, we drop indices $i$ and $h$ indicating the discretizations in time and space. 
Also, we reduce the number of model parameters by non-dimensionalizing the governing equations. All simulations are started with an initially spherical surface, the radius $R$ of which we choose as characteristic length-scale. Moreover, we set the diffusion time  $\tau_D=\frac{ R^2}{D_c}$ as characteristic time-scale, and the characteristic concentration $\frac{k_\mathrm{off}}{k_\mathrm{on} c_\mathrm{0}}$. 
Using $\eta_\mathrm{b}$ as characteristic viscosity, the non-dimensionalization of Eq.~\eqref{eq:force} results in dimensionless values for surface mass density $\hat{\rho}_\Gamma=\frac{D_c \rho_\mathrm{\Gamma}}{\eta_\mathrm{b}}$, volume correction parameter $\hat{\alpha}=\frac{R^3\alpha}{D_c \eta_\mathrm{b}}$, P\'{e}clet number $Pe=\frac{\xi R^2}{D_c \eta_\mathrm{b}}$, grid spacing $\hat{h}=\frac{h}{R}$, time $\hat{t}=\frac{t}{\tau_\mathrm{D}}$, time-step $\hat{k}=\frac{k}{\tau_\mathrm{D}}$, relaxation times $\hat{\tau}_\mathrm{b,s}=\frac{\tau_\mathrm{b,s}}{\tau_\mathrm{D}}$, stresses $\hat{S} = \frac{S\tau_D}{\eta_\mathrm{b}}$,  and viscosity ratio $\nu=\frac{\eta_\mathrm{s}}{\eta_\mathrm{b}}$.

\subsection{Validation 1: Growth and shrinkage of a sphere}
\label{subsec:Stretching_sphere}

As a first validation, we simplify our model by including only dilational stress ($\tr\mathbf{S}$), in the absence of shear stress ($\bar{\mathbf{S}} = 0$, $\nu=0$). We start with a unit sphere $\Gamma$ centered at the origin and solve the system for a prescribed velocity field $\hat{\boldsymbol{v}} = \sin(\hat{t}) \hat{\boldsymbol{x}}$, 
corresponding to a periodic inflation/deflation of the sphere, similarly to Kinkelder et. al \cite{Kinkelder2021}. The initial condition corresponds to an unstretched sphere, with $\hat{\mathbf{S}}=0$. 
For this reduced model, the development of the dilational stress is spatially constant and leads to the ODE 
\begin{equation}
\tr \hat{\mathbf{S}} = 4 \sin{(\hat{t})} + \hat{\tau}_\mathrm{b} (2 \tr\hat{\mathbf{S}} \sin{(\hat{t})} - \partial_{\mathrm{\hat{t}}} \tr \hat{\mathbf{S}})~,
\end{equation}
which can be solved analytically.

\begin{center}
\begin{figure}
  \includegraphics[width = 1\textwidth]{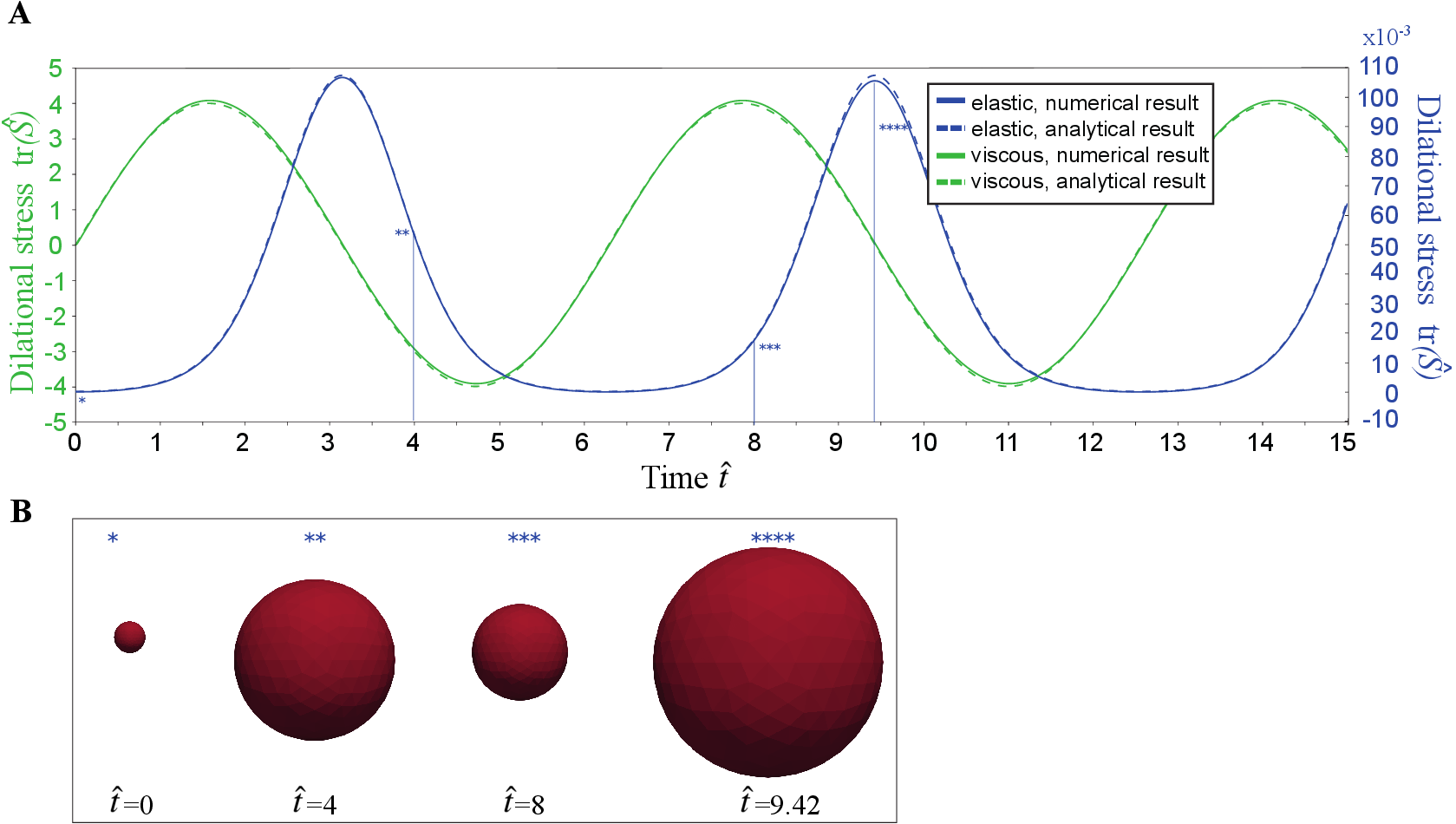}
  \caption{\textbf{A.}The dilational stress $\tr\hat{\mathbf{S}}$ for the purely viscous (green curves) and the elastic case (blue curves). The analytical solutions (dashed lines) are given by equations (\ref{eq:viscous}) and (\ref{eq:elastic}). For the viscous case ($\nu=0$, $\hat{\tau}_\mathrm{b}=10^{-2}$), the numerical result (green, solid line) fits very well with the analytical result (green, dashed line). For the elastic case ($\nu=0$ and $\hat{\tau}_\mathrm{b}=10^{3}$) the numerical result (blue,solid line) deviates more from the analytical result (blue, dashed line) with increasing time, corresponding to stress dissipation. \textbf{B.} Inflation and deflation of the sphere for the elastic case, shown at different simulation times ($*$-symbols mark the corresponding time point in panel A). \newline 
Simulation are performed with $\hat{h} = 0.2$, $\hat{k} = 10^{-3}$, $\hat{\rho}_\mathrm{\Gamma}=0.001$, $\hat{\alpha}=0$.}
  \label{fig:growing_sphere}
\end{figure}
\end{center}

We validate the dilational stress separately for the viscous and elastic limit case.
\begin{itemize}
\item[(1)] In the viscous case (i.e. $\hat{\tau}_\mathrm{b} \ll 1$), the analytical solution is then given by
\begin{equation}
\tr \hat{\mathbf{S}} = 4 \sin(\hat{t}) ~, 
\label{eq:viscous}
\end{equation} 
which is shown in Fig. \ref{fig:growing_sphere} (green, solid line). 

\item[(2)] In the elastic limit case, we choose 
$\tau_\mathrm{b} \gg 1 $, whereupon  the in-plane dilational stress should  approximate that of a neo-Hookean shell model. 
The analytical solution reads then 

\begin{equation}
\tr \hat{\mathbf{S}} = \frac{2}{\hat{\tau}_\mathrm{b}} \left( \exp \left( 2 (1-\cos(\hat{t})) \right) - 1 \right) ~,
\label{eq:elastic}
\end{equation}
see Fig. \ref{fig:growing_sphere} (blue, solid line).
\end{itemize}

Figure~\ref{fig:growing_sphere} compares analytical and numerical results for both limiting cases. Consistent with findings by Kinkelder et al.~\cite{Kinkelder2021}, we observe excellent agreement; in fact, our numerical results show even closer alignment with the analytical reference than those reported in~\cite{Kinkelder2021}. In the elastic regime, deviations grow slightly over longer simulation times due to residual viscous stress dissipation. 
We therefore conclude that our model yields stable and accurate results when the system is governed solely by dilational stress. 

\subsection{Validation 2: Shear stress by critical P\'{e}clet number}
\label{subsec:Peclet_number}
After the validation of the dilation stress, we validate the shear stress part of the viscoelastic model. 
To do so, we simulate the spontaneous pattern formation of the viscoelastic surface. It has been shown previously, that the viscous limit of this pattern forming system depends sensitively on the shear stress of the system \cite{Mietke2019, Wohlgemuth2023}. \newline 
The value of interest here is the P\'{e}clet number, which is the ratio between the active advection rate and the diffusion rate within the system. 
If the surface activity $\xi$ and thus the P\'{e}clet number is large enough, the system is able to form concentration patterns by a mechano-chemical feedback mechanism\cite{Wittwer2023}. 
The critical P\'{e}clet number, above which a surface with initial concentration $c_0$ forms stationary patterns, is given by \cite{Mietke2019}
\begin{equation}
Pe_l^*=\frac{1}{c_0 \partial_c f(c_0 )} \left( 1+ \frac{\tau_D k_\mathrm{{off}}}{ l (l+1)} \right) \left( l(l+1) + \nu ((l-1)(l+2)) \right)~,
\label{eq:critical_peclet}
\end{equation}
Here, $l$ denotes the mode number of axisymmetric spherical harmonics corresponding to the observed pattern. The lower-order modes, specifically $l=1$ (corresponding to a polar pattern) and $l=2$ (representing a ring pattern), are the most unstable ones and of particular biological relevance.

\begin{center}
\begin{figure}
  \includegraphics[width = 1\textwidth]{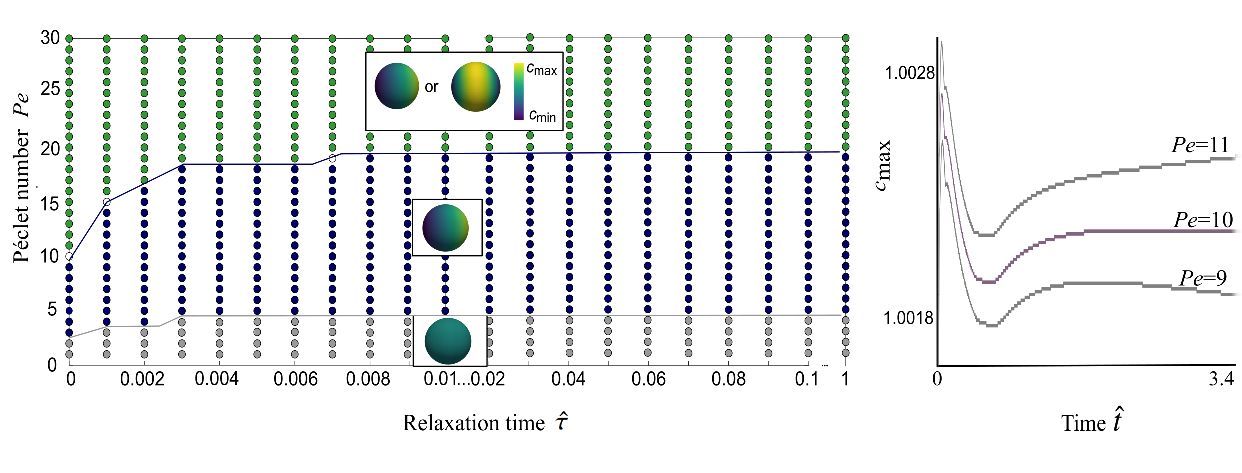}
  \caption{ 
  \textbf{Left}:
  Phase diagram of growing patterns in dependence of  P\'{e}clet number $Pe$ and relaxation time $\hat{\tau}=\hat{\tau}_\mathrm{b}=\hat{\tau}_\mathrm{s}$. In the viscous limit ($\hat{\tau}=0$), the correct critical P\'{e}clet numbers are recovered.
  \textbf{Right}:
  Evolution of peak concentration arising from an initially nearly uniform concentration field, perturbed with a slight $l=2$ mode. Depending on the P\'{e}clet number, the pattern grows (\( Pe = 11 \)), saturates (\( Pe = 10\)), or decreases (\( Pe = 9\)). The transition at $Pe=10$ agrees with the critical P\'{e}clet number $Pe_2^*$ predicted by linear stability analysis. 
  Numerical parameters: $\hat{h} =0.1$,$\hat{k}=10^{-4}$, $\nu=1$, $\hat{\rho}_\Gamma=0.001$, $\hat{\alpha}=10^4$, $\tau_D k_{\mathrm{off}}=0$.
  }
  \label{fig:Peclet_number}
\end{figure}
\end{center}

To validate our numerical model, we investigate pattern formation across a range of Péclet numbers. 
We initialize the concentration field at a baseline value \( c_0 \), perturbed by a small amplitude pattern corresponding to either the \( l = 1 \) or \( l = 2 \) mode, and observe whether these modes amplify or decay over time. 
Figure~\ref{fig:Peclet_number}(right) illustrates the behavior of the \( l = 2 \) mode in the viscous regime, defined by \( \hat{\tau}_\mathrm{b} = \hat{\tau}_\mathrm{s} = 0 \). 
As the Péclet number varies, the mode either grows (\( Pe = 11 \)), saturates (\( Pe = 10 \)), or decays (\( Pe = 9 \)). 
The observed transition at \( Pe = 10 \) is consistent with the critical Péclet number \( Pe_2^* = 10 \) predicted by linear stability analysis.

We extend these results by constructing a phase diagram (Figure~\ref{fig:Peclet_number} left) covering a broader range of Péclet numbers and relaxation times. 
In the viscous regime (\( \hat{\tau}_\mathrm{b} = \hat{\tau}_\mathrm{s} = 0 \)), the model recovers the critical threshold \( Pe_1^* = 2 \). 
Below this value, no pattern formation is observed, as \( Pe_l \geq 2 \) for all \( l \). 
For \( 3 \leq Pe < 10 \), only the first mode (\( l = 1 \)) grows, whereas for \( Pe > 10 \), both \( l = 1 \) and \( l = 2 \) modes exhibit growth.

In the viscoelastic regime ($\hat{\tau}>0$), analytical solutions are not available. Our phase diagram thus offers a first insight into active pattern formation on viscoelastic surfaces. As relaxation times increase, greater activity ($Pe$) is required to initiate pattern formation. This trend saturates at Péclet number of approximately twice the value observed in the purely viscous regime.

\subsection{Validation 3: Convergence study}
\label{subsec:convergence_study}
To verify the numerical scheme further, we conduct convergence tests for the full numerical viscoelastic model. We consider a unit sphere as initial surface, with initial concentration given by $c(\theta) = -\frac{3}{2} \cos^2\left(\theta +\frac{1}{2}\right)$, where $\theta$ is the azimuthal angle around the z axis with an angle of 0 corresponding to the x axis. We run simulation for a specific time interval until $\hat{t} = 6\times10^{-3}$.
This is done for four levels of refinement with mesh width $h = 0.132, 0.092, 0.068, 0.049$. The timestep size for the largest $h$ is $\hat{k}=10^{-3}$. The timestep sizes for smaller $h$ are chosen to fulfill the time step restriction $h^3 \sim \hat{k}$, ensuring that the spatial error dominates the order of convergence. The solution for the smallest $h$ is considered as reference solution. We compute the relative errors with respect to the $L^2$-norm in space and the $L^\infty$-norm in time, i.e. for some quantity $q$, we compute
\begin{align*}
    e_q = \left\|\frac{\|\hat{q}_h - \hat{q}\|_{L^2(\Gamma)}}{\|\hat{q}\|_{L^2(\Gamma)}}\right\|_{L^\infty([0,\hat{t}])},
\end{align*}
where $\hat{q}_h$ is the discrete dimensionless solution with grid width $h$ and $\hat{q}$ is the dimensionless reference solution.
We consider the errors of the concentration field $\hat{c}$, the velocity field $\hat{\boldsymbol{v}}$, the total curvature $\hat{H}$, the parametrization $\hat{\mathbf{x}}$, the normal $\hat{\boldsymbol{n}}$ and the enclosed volume $\hat{V}$.


\begin{figure}
    \centering
    \includegraphics[width = 0.75\textwidth]{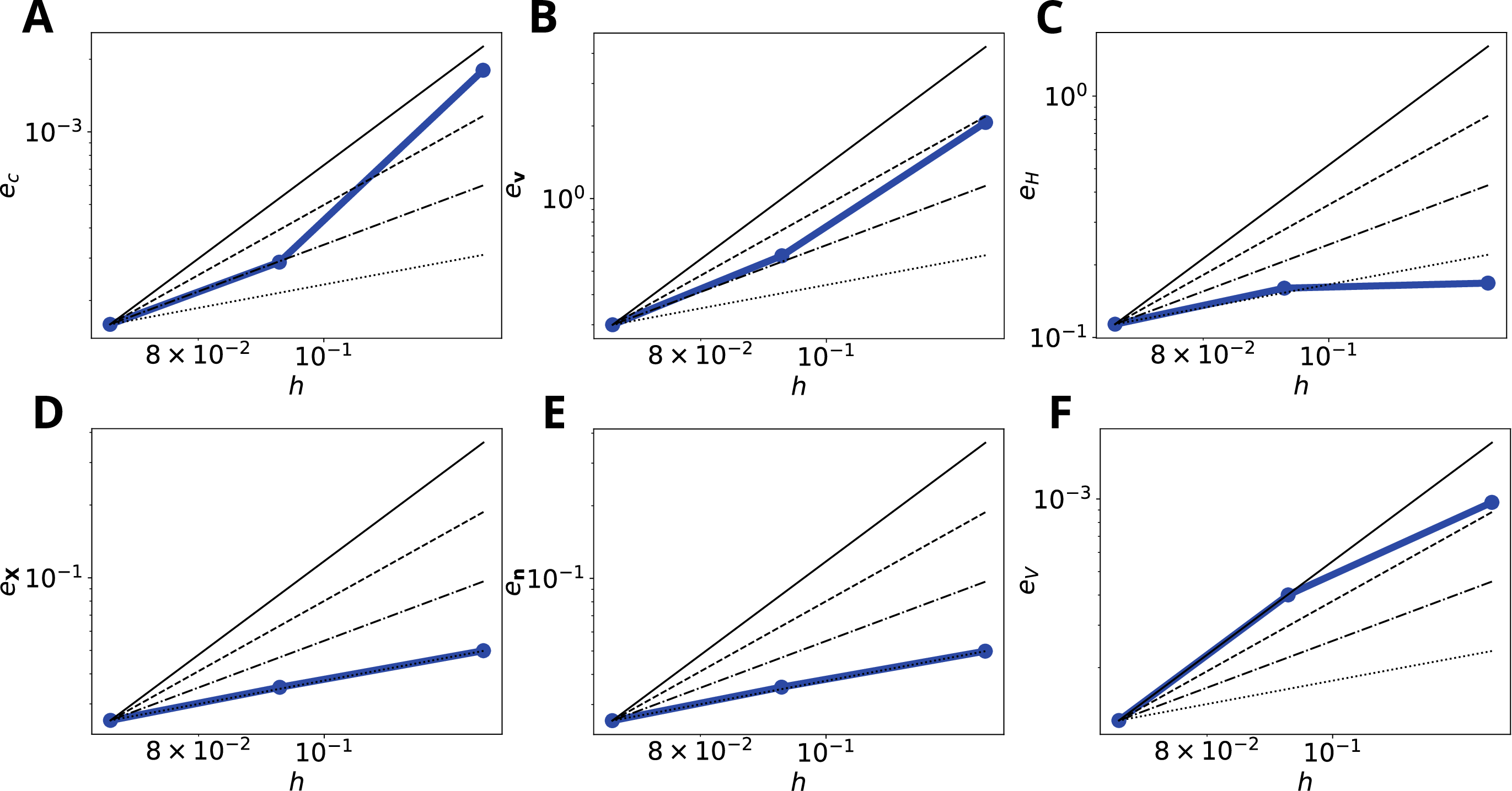}
    \caption{Convergence study for viscoelastic model with respect to the mesh size $h^3 \sim k$. Shown in blue are the errors for the concentration $\hat{c}$ (A), the velocity $\hat{\boldsymbol{v}}$ (B), the mean curvature $\hat{H}$ (C), the parametrization $\hat{\mathbf{x}}$ (D), the normal $\hat{\boldsymbol{n}}$ (E), and the enclosed volume $\hat{V}$ (F). Convergence orders of 1, 2, 3 and 4 are indicated by dotted, dash-dotted, dashed and solid lines (black), respectively.}
    \label{fig:convergence}
\end{figure}

The errors are shown in Fig. \ref{fig:convergence} for parameters $\hat\tau_b = \hat\tau_s = 0.001$, $Pe = 80$, $\hat\rho_\Gamma=0.0001$, $\nu = 1$, $\hat\alpha=10^5$. All quantities converge for decreasing mesh width. Due to the lack of theoretical studies on the convergence properties of this model, we refrain from making any assertions about its convergence rates.

\subsection{Dynamics of ring pattern formation on the actin cortex}
\label{sec:Space_and_time_dependent_attachment_rate}

\begin{center}
\begin{figure}
  \includegraphics[width = 0.9\textwidth]{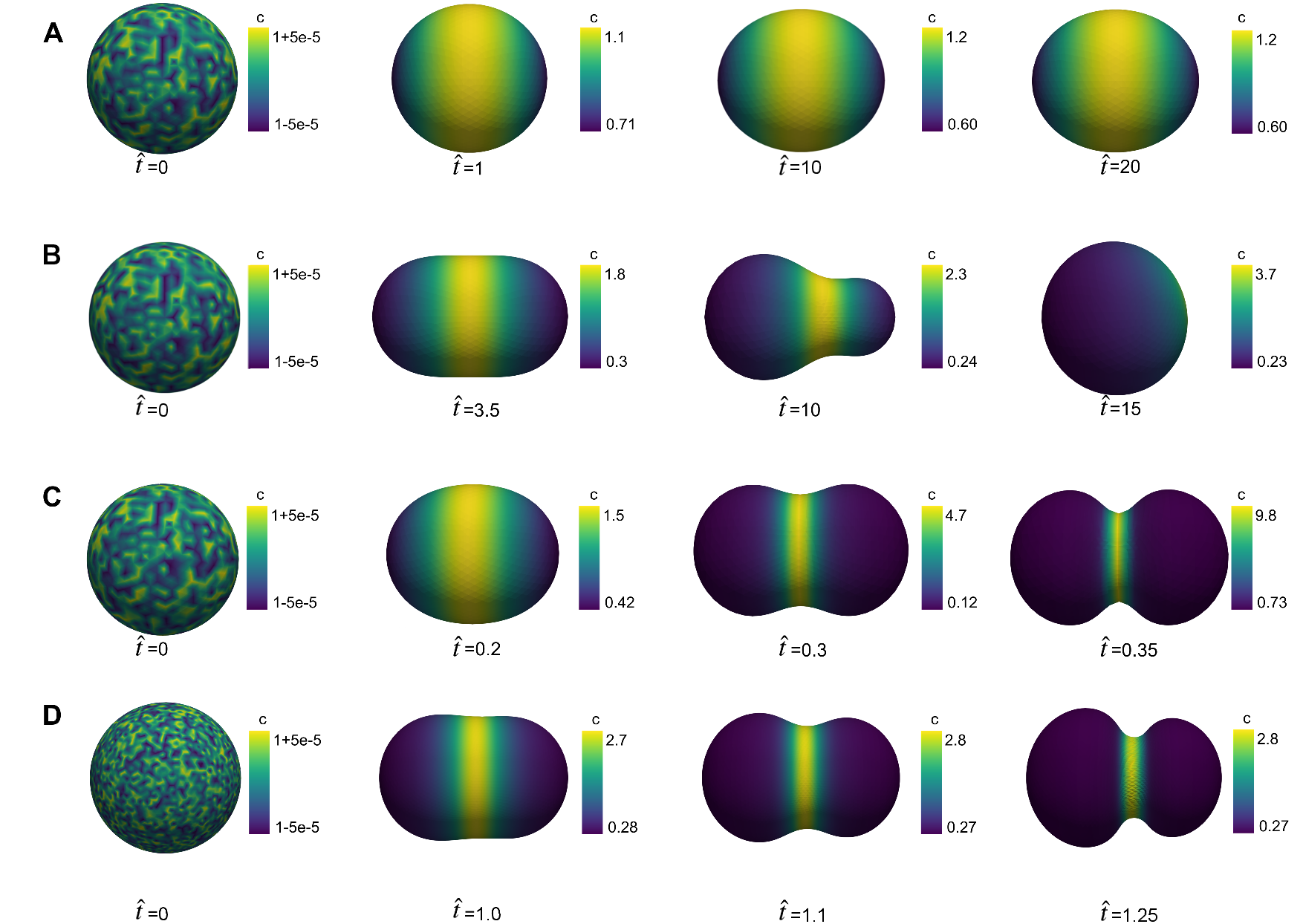}
  \caption{\textbf{A. Stable state} below critical Péclet number $Pe=5$, here for the viscous case $\nu=1$, $\hat{\tau}_\mathrm{b}=\hat{\tau}_\mathrm{s}=0$. \textbf{B. Slipping ring} above critical Péclet number $Pe=20$ for the viscous case, with $\nu=1$, $\hat{\tau}_\mathrm{b}=\hat{\tau}_\mathrm{s}=0$, $\beta_\mathrm{0}=0.5$. \textbf{C. Onset of symmetric cell division dynamics} for the viscoelastic case, with $Pe=20$, $\nu=1$, $\hat{\tau}_\mathrm{b}=\hat{\tau}_\mathrm{s}=0.01$, $\beta_0=0.5$. \textbf{D. Onset of asymmetric cell division dynamics} for the viscoelastic case, with $Pe=20$, $\nu=1$, $\hat{\tau}_\mathrm{b}=\hat{\tau}_\mathrm{s}=0.005$, $\beta_0=0.01$. All other simulation parameters were chosen as follows: $\hat{h}=0.1$ (panel A, B and C), $\hat{h}=0.05$ (panel D), $\hat{k} =10^{-3}$, $\hat{\rho}_\Gamma=0.001$, $\hat{\alpha}=10^5$.}
  \label{fig:patterns}
\end{figure}
\end{center}

During the onset of cell division, a ring enriched in active force-generating molecules assembles in the division plane within the cell cortex and gradually constricts the cell, potentially aided by the viscoelastic response of the cortical surface, as described here.  Until now, such mechanisms have only been simulated for purely viscous surfaces. In these systems, the ring forms in the division plane but subsequently slips toward one side of the sphere \cite{Kinkelder2025}, a behavior we refer to here as \textit{slipping ring dynamics}, see also Fig.~\ref{fig:patterns}B. 

Similar to Mietke \textit{et al.} \cite{Mietke2019}, we add a space dependent attachment rate for the concentration (see equation (\ref{eq:concentration})), to account for the orientation of the mitotic spindle in the biological system and the respective inhomogeneous biochemical signaling. This new attachment rate is defined as follows:
\begin{equation}
    \tilde{k}_\mathrm{on}(\theta)= k_{\mathrm{on}} \left(1+ \beta_\mathrm{0} (1- 3 \cos^2(\theta))  \right) 
    \label{eq: k_on of theta}
\end{equation}
To ensure that this rate acts as an attachment rate, we choose $\beta_\mathrm{0}$ in the interval $[0,1/2]$ so that $k_{\mathrm{on}}(\theta)>0$, with the azimuthal angle $\theta$. Here, for a point $\mathbf{x}$ on the surface, we consider the corresponding azimuthal angle of the point projected to the unit sphere $\frac{\mathbf{x}}{|\mathbf{x}|}$ and the angle is oriented in the same way as above. With this modification we aim to analyze possible dynamics of the active surface, and the influence of the viscoelasticity. 

We initialized simulations on a spherical surface, starting with a slightly perturbed homogeneous concentration field ($\hat{c}=1\pm5\cdot 10^{-4}$). 
First, using a P\'{e}clet number smaller than the ciritcal P\'{e}clet number $Pe<Pe_1^*$, the system shows a stable wide ring, as expected due to the attachment rate, because all spontaneously forming modes must relax to a stable state, see Fig. \ref{fig:patterns}A. 
In this scenario the sphere deforms to become gradually more prolate. 
In contrast, in the regime of pattern formation, we observe three possible patterns. For the viscous case ($\hat{\tau}_\mathrm{b}=\hat{\tau}_\mathrm{s}=0$),  a slipping ring dynamics occurs, see Fig. \ref{fig:patterns}B, in agreement with previous literature \cite{Kinkelder2025, Wittwer2023, Mietke2019}. 
The corresponding stationary state is characterized by a single pole of elevated concentration. Notably, despite continuous attachment along the ring via $k_{\rm on}$, the incorporated material is persistently advected toward the pole, resulting in its accumulation.

For a viscoelastic surface, the dynamics changes: The ring grows and can be stabilized in the middle, which leads to a cell dynamics resembling the onset of symmetric cell division, see Fig. \ref{fig:patterns}C. This behavior indicates that viscoelasticity can stabilize the cytokinetic ring. 
Additionally, a third possible pattern was found in the viscoelastic case, for smaller attachment rates, see Fig. \ref{fig:patterns}D. The cell gradually constricts, while the ring slips to one side of the sphere. We propose that such dynamics could contribute to the phenomenon of asymmetric cell division in biological systems.

To screen the dynamical behavior of the viscoelastic model, multiple simulations were performed, in which the attachment rate was scaled by different values for $\beta_\mathrm{0}$ and for different relaxation times $\hat{\tau}=\hat{\tau}_\mathrm{b}=\hat{\tau}_\mathrm{s}$, see Fig. 6. All simulations were performed with the same randomized initial conditions and $Pe=20$ , as before. 
In the absence of a space dependent  attachment rate ($\beta_\mathrm{0}=0$), the first spherical harmonics mode is dominant, which leads to a concentration focus on one side of the sphere, see Fig. 6, (orange regime). For a space dependent attachment rate $\beta_\mathrm{0}>0$, we observed slipping ring dynamics for small relaxation times (green regime), onset of symmetric cell division for larger relaxation time (purple regime) or onset of asymmetric cell division (cyan regime). 

\begin{center}
\begin{figure}
  \includegraphics[width = 1\textwidth]{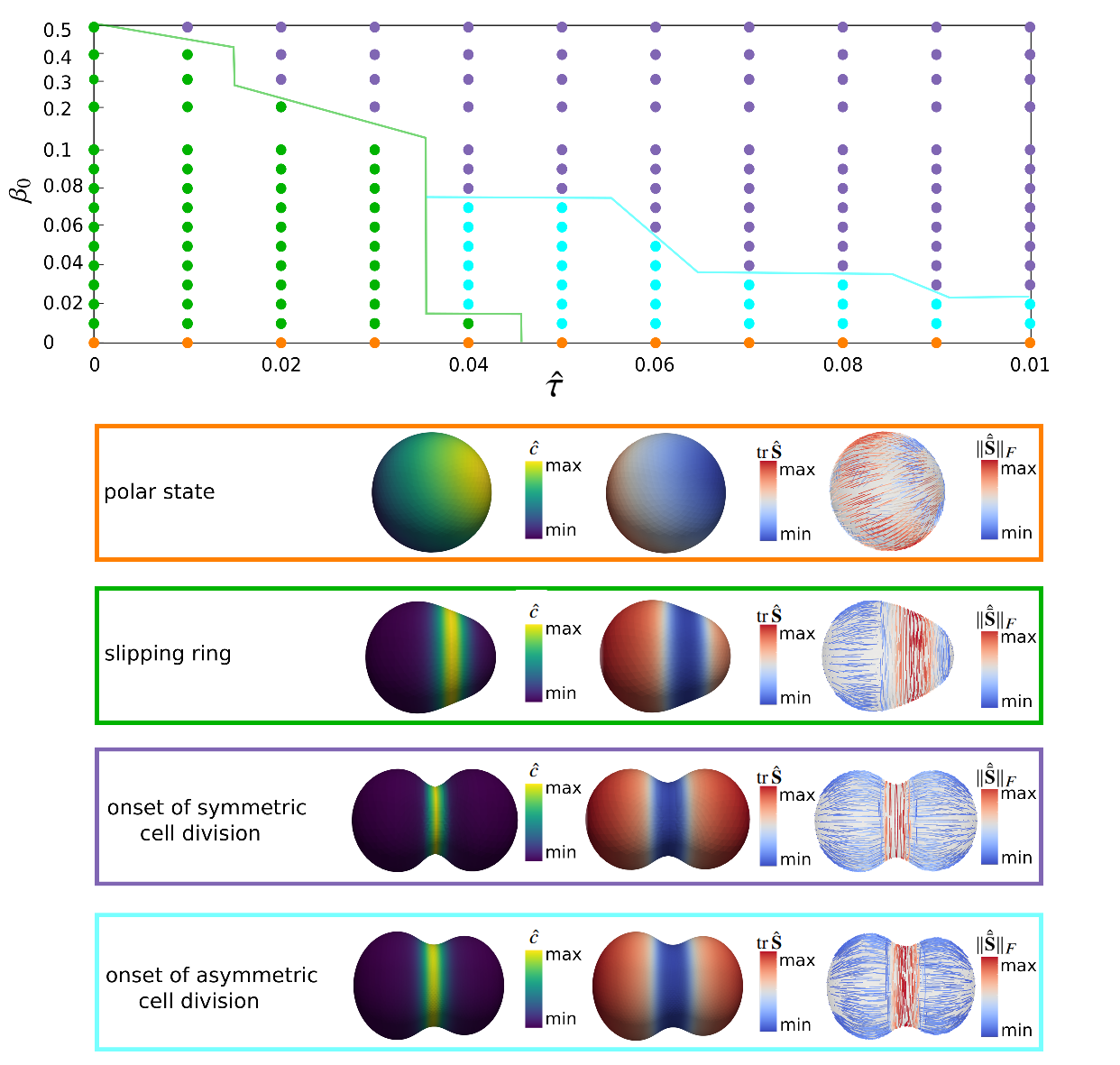}
  \caption{\textbf{Phase diagram} of pattern formation of the active viscoelastic surface. In the limit, of a viscous surface without space dependent attachement rate ($\beta_0=0$), the first mode is dominant and the concentration assembles on one side of the sphere, which we declare as polar state (orange). For a space dependent attachment rate ($\beta_0>0$), three possible behaviors have been observed: slipping ring dynamics (green), onset of symmetric cell division (purple) and onset of asymmetric cell division while the ring slips (cyan). 
  Bottom rectangles show exemplary snapshots of the four regimes highlighting concentration $\hat c$, dilational stress $\tr \hat{\mathbf{S}}$, and the eigenvector corresponding to the largest eigenvalue (by magnitude) of $\hat{\bar{\mathbf{S}}}$ colored by $\| \hat{\bar{\mathbf{S}}}\|_\text{F}$, respectively.  \\
  All simulations were performed with: $Pe=20$, $\nu=1$, $\hat{h}=0.1$, $\hat{k} =10^{-3}$,  $\hat{\rho}_\mathrm{\Gamma}=0.001$, $\hat{\alpha}=10^5$.}
\label{fig:pd_possible_patterns}
\end{figure}
\end{center}

\subsection{Dynamics of ring stabilization on ellipsoidal surfaces}
\label{C_elegans}
{
As a final example, we demonstrate the model’s ability to capture three-dimensional effects in active surface pattern formation. A particularly intriguing phenomenon is the alignment of the cytokinetic ring with the long axis of non-spherical embryonic cells during division. This alignment, driven by the self-organizing dynamics of the emerging ring pattern, has been observed in several embryos, including \textit{C. elegans} (\textit{Caenorhabditis elegans}) and mouse \cite{middelkoop2024cytokinetic}. Remarkably, this behavior persists even when the mitotic spindle -- initially responsible for triggering ring formation -- is misaligned with the cell’s long axis. In such cases, the ring pattern undergoes a self-organized rotation to achieve alignment, illustrating the robustness of this mechanism.

To investigate this scenario, we simulate an ellipsoidal surface with the axis lengths 1:1:2, to cover the shape of \textit{C. elegans}, see Fig.~\ref{Fig. 7}. 
In the biological system, the change of the surface shape is restricted by a surrounding egg shell. 
To describe this constriction, we suppress movement in the direction normal to the surface by changing the weak formula of the force balance \eqref{eq:weak_force_balance} to:
\begin{align}
\int \left( \varepsilon \left( \textbf{v}^{i+1} \cdot \boldsymbol{n}_h^i \right) \boldsymbol{n}_h^i \right) \cdot \textbf{u} ~d \Gamma &= \int_{\Gamma_h} \left(\rho_\Gamma \left(\frac{\boldsymbol{v}_h^{i+1}-\boldsymbol{v}_h^i}{k} + \nabla_{\boldsymbol{w}_h^i}\boldsymbol{v}^{i+1} \right) + q\boldsymbol{n}_h^i \right) \cdot \boldsymbol{u} ~{\textrm d}\Gamma  \notag \\
&- \int_{\Gamma_h} \xi \left(f'(c_h^i)\nabla_\Gamma c_h^{i+1}  \right) \cdot \boldsymbol{u} ~{\textrm d}\Gamma \notag \\
&- \int_{\Gamma_h} {\textbf{f}}_h^{i+1} \cdot {\boldsymbol{u}} ~{\textrm d}\Gamma ~.
\label{eq:weak_force_balance_c_elegans}
\end{align}
The new term on the left-hand side is an effective force which penalizes any movement in normal direction for suitably large $\varepsilon$. 

We prescribe a misaligned ring pattern which is rotated by $20$\textdegree (see Fig. \ref{Fig. 7}) ~by defining the initial  concentration as
\begin{equation}
\hat{c} = 1 \pm 0.01 \exp\left( -\left( \frac{\cos\left( \theta+ 20^\circ \right)}{0.4} \right)^2 \right), 
\end{equation}
where $\theta$ is again the azimuthal angle around the z-axis. The attachement rate $k_\mathrm{on}$ is chosen as a constant, i.e. not in the spatially varying form from Eq.~\eqref{eq: k_on of theta}.

Our simulation reveals a two-stage process for ring self-correction: a misaligned ring robustly rotates to the central axis (Fig.~\ref{Fig. 7}). 
Initially, two local maxima in concentration emerge and migrate toward the tips of the embryo (top right and bottom left at $\hat{t}=0.6$). 
Subsequently, as the concentration values in these maxima continuously grow, their positional dynamics inverts, and they move toward the center of the long axis. The rotation concludes when these concentration maxima reassemble into a stable ring ($\hat{t}=1.9$).
Our simulations therefore highlights the robustness of the active surface model's ring positioning mechanism, matching predictions for the biological system \cite{middelkoop2024cytokinetic}.

\begin{figure}
  \centering
  \includegraphics[width = 1\textwidth]{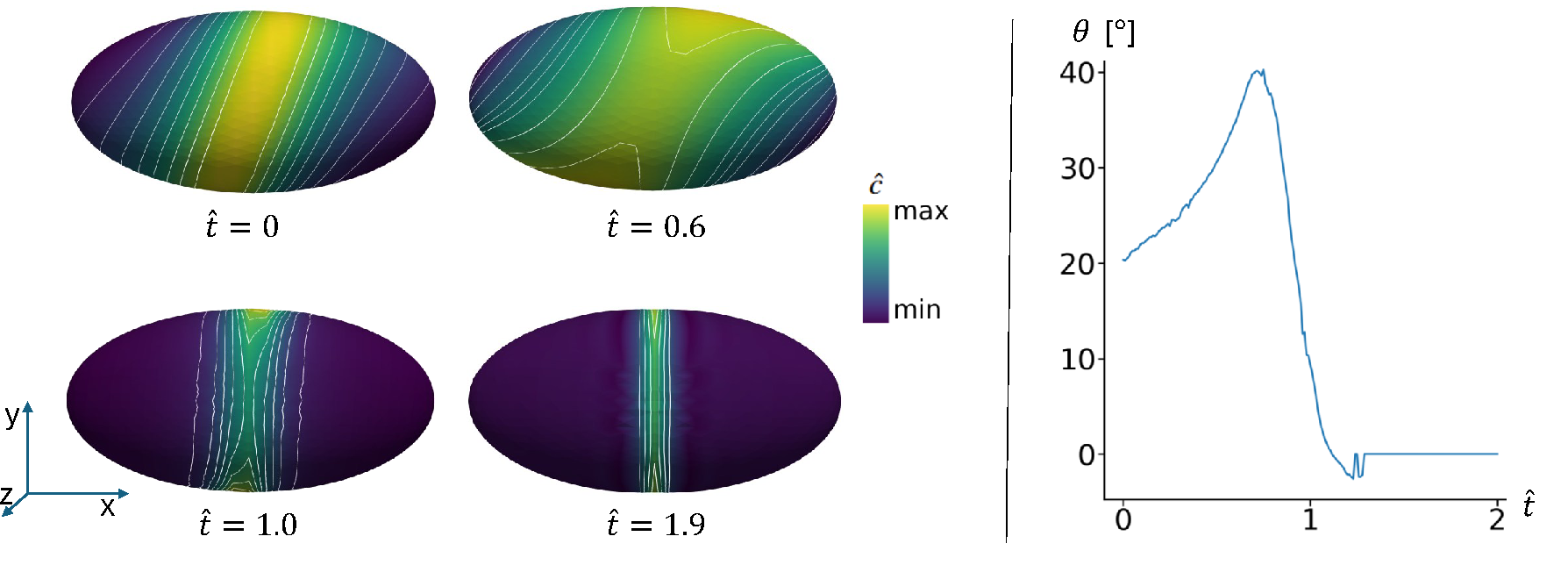}
  \caption{\textbf{Stabilization of the contractile ring.} Simulation of a fixed ellipsoidal surface with axis ratio 1:1:2 to cover the shape of the \textit{C. elegans} embryo. 
  \textbf{Left:} The initally concentration ring is rotated by $20^\circ$ around the z-axis  ($\hat{t}=0$). The ring rotates toward the center of the long axis ($\hat{t}=1.9$) via two transiently emerging concentration spots (visible at $\hat{t}=0.6$ and $\hat{t}=1.0$), while the peak concentration  grows monotonically over time. White lines are level sets of $\hat{c}$ as guide for the eye.   
  \textbf{Right:} Angle between the yz-plane and the vector from ellipsoid barycenter to the highest concentration point over time. The initially imposed ring inclination of $20$\textdegree ~first increases and then decreases to zero. 
  Parameters:  $Pe=30$, $\nu=1$, $\hat{\tau}_\mathrm{s}=\hat{\tau}_\mathrm{b}=0.003$, $\hat{k} =10^{-3}$,  $\hat{\rho}_\mathrm{\Gamma}=0.001$, $\varepsilon =10^3, \hat{h}=0.1.$ 
  }
  \label{Fig. 7}
\end{figure}
}

\section{Discussion}
\label{sec:Discussion}

In this work, we introduced a numerical framework for simulating active viscoelastic surfaces. Our approach relies on a monolithic implementation, where all surface properties, are solved simultaneously within a single coupled system. The computational grid is then updated in a subsequent step using the Arbitrary Lagrangian-Eulerian technique. We demonstrated that this strategy is effective and show numerical convergence and excellent agreement between numerical and analytical solution for the limit cases of purely viscous and purely elastic surfaces under dilations. 

We applied the model to simulate a key biological system — the active actin cortex surface in dividing eukaryotic cells. To capture enhanced activation of cortex-associated force-generating molecules in the midzone of the mitotic spindle, we introduced a spatially dependent attachment rate with increased attachment in the center \cite{Glotzer2013}.  
Our simulations show that for a sufficiently high surface activity, a viscous cortical surface forms a ring in the center, which slips to one side, as predicted by previous models \cite{Mietke2019, Kinkelder2025, Wittwer2023}. For a viscoelastic surface, the ring stabilizes for sufficiently large relaxation times allowing for successful cell constriction. This cell constriction can either be symmetric or asymmetric with lower attachment rate bias to the midzone favoring asymmetric constriction. 

As the dimensionless relaxation times used in our simulations are consistent with biological values \cite{Fischer2016} (in the range of $1-100$s), our study identifies finite viscoelastic relaxation times as a previously missing model ingredient required to stabilize emergent contractile rings in a minimal model of active actin cortex surfaces. In doing so, we address a long-standing question of how such rings can robustly support cell division. Controlled asymmetric cell divisions are an important requirement for the development of healthy multicellular organisms \cite{Sunchu2020}. Our model provides initial insights into how the cortical system might regulate the choice between asymmetric and symmetric cell divisions. 

Moreover, we illustrate how the model can be used to simulate the self-correction dynamics of misaligned rings on ellipsoidal viscoelastic surfaces.
In conclusion, we expect that our new numerical method for the simulation of active viscoelastic surfaces may serve as a versatile tool to more deeply investigate the physical principles that govern cytokinesis and other self-deformation processes of active biological surfaces. 


\begin{ack}
We thank Marcel Mokbel for the fruitful discussions and AMDiS support. All authors gratefully acknowledge computing time on the
high-performance computer at the NHR Center at TU Dresden. This center is jointly supported by the Federal Ministry of Education and Research and the state governments participating in the NHR (www.nhr-verein.de/unsere-partner)
\end{ack}

\begin{funding}
This work was supported by DFG via research unit FOR 3013 (grant 417223351 to SA, EFF and AV).
This research was supported in part by grant NSF PHY-2309135 and the Gordon and Betty Moore Foundation Grant No. 2919.02 to the Kavli Institute for Theoretical Physics (KITP) in Santa Barbara, US, as well as the Swedish Research Council under grant no. 2021-06594 to the Institut Mittag-Leffler in Djursholm, Sweden.
\end{funding}


\end{document}